\documentclass[aps,twocolumn,amsmath,letterpaper]{revtex4}
\usepackage{amssymb}
\usepackage{graphicx}
\usepackage{array}
\usepackage{hhline}
\usepackage{longtable}
\cleardoublepage

\renewcommand{\thetable}{\Roman{table}} \thetable

\begin{document}

\title{The Critical Properties of Two-dimensional Oscillator Arrays}

\author{Gabriele Migliorini}
\affiliation{ The Neural Computing Research Group\\ School of
Engineering and Applied Sciences, Aston University\\ Birmingham B4
7ET, United Kingdom.}

\begin{abstract}

We present a renormalization group study of two dimensional arrays 
of oscillators, with dissipative, short range interactions. We 
consider the case of non-identical oscillators, with distributed 
intrinsic frequencies within the array and study the 
steady-state properties of the system. 
In two dimensions no macroscopic mutual entrainment is found but, 
for identical oscillators, critical behavior of the 
Berezinskii-Kosterlitz-Thouless type is shown to be present.
We then discuss the stability of (BKT) order in the physical case of 
distributed quenched random frequencies.
In order to do that, we show how the steady-state dynamical properties 
of the two dimensional array of non-identical oscillators are related 
to the equilibrium properties of the $XY$ model with quenched randomness, 
that has been already studied in the past. We propose 
a novel set of recursion relations to study this system within 
the Migdal Kadanoff renormalization group scheme, by mean of the discrete 
clock-state formulation. We compute the phase diagram in the presence 
of random dissipative coupling, at finite values of the clock state 
parameter. 
Possible experimental applications in two dimensional arrays of 
microelectromechanical oscillators are briefly suggested. \\


\end{abstract}
\maketitle
\def\s{\rule{0in}{0.28in}}
\section{Introduction}

The study of the dynamical properties of large arrays of 
self-sustained oscillators 
with distributed intrinsic frequencies is an interesting 
problem bridging non-equilibrium statistical physics and non-linear 
science. Large populations of interacting, self-sustained oscillators 
are known to model a great variety of biological systems and several 
progresses have been made in the last decades \cite{zero,zerob,zeroc}. 
If dense arrays of coupled self-oscillators, close to an oscillatory 
instability of the Hopf type, are expected to present interesting 
critical behavior, the analysis of such systems in the 
low dimensional case, i.e. when the range of the interactions is short, 
has been quite a challenging one and very little attempts have been
made in this direction.  
The mean-field Kuramoto model of phase oscillators represents a very 
special case where one can predict in a clear way the macroscopic mutual 
entrainment (MME) properties of the system, but attempts to generalize such
results to the low dimensional case are usually difficult and the
majority of the studies of oscillator arrays have been devoted to 
the $all~to~all$ case, and to networks with relatively high
degree of connectivity. Meanwhile, the case of short range interactions 
is a very interesting one and have a lot of potential applications, 
in particular considering the case of dissipatively and/or reactively 
coupled arrays of mechanical oscillators at micro and nano scales.

The oscillator lattice problem, namely the finite-dimensional, nearest 
neighbour version of the Kuramoto model, has been attacked in the past 
by several authors \cite{one,two}. 
The main conclusions are that the synchronization 
properties one normally finds in the mean-field case are drastically 
changed by the short range nature of the connectivity, and that no
synchronization is expected in two dimensions \cite{one,two,twob}. 
An interesting analysis 
based on real space renormalization group theory \cite{three} suggests 
an explicit expression for the lower critical dimension, below which one 
do not expect macroscopic mutual entrainment, the latter being the dynamical
analogue of ferromagnetic order.
A related situation occurs within the equilibrium statistical physics of 
spin systems with continuous degree of freedom of the $XY$ type, 
as already suggested in \cite{one}. 
In the statistical mechanics of continuous spin models of the mean-field
type the system has a spontaneous magnetization and the out-of-equilibrium  
behavior is a very interesting and open problem \cite{threeb}. These 
models relate closely to the Kuramoto model, where the additional difficulty 
of the intrinsic, quenched frequency distribution has to be taken into
account, and where the system does not reach equilibrium but is self-driven
in an out-of-equilibrium steady state regime.

Both the dynamical and equilibrium properties of continuous spin models 
change importantly when short ranged, nearest neighbour interactions 
are considered. Rather than ordinary symmetry breaking, in the 
two-dimensional $XY$ model algebraic order (AO) occurs. 
Namely the system has infinite correlation length, but no 
magnetization at finite temperature \cite{threec,threed,threee}.

When considering the case of low dimensional arrays of coupled
oscillators, the problem is further complicated by the fact that 
the system is driven out-of-equilibrium, so that the standard tools
of equilibrium statistical mechanics cannot apparently be exploited.
On the other side, if the analogy between arrays of identical oscillators and the 
statistical physics of $XY$ type of models with continuous degrees of
freedom is a well known fact and attempts to relate the dynamical
properties of low-dimensional arrays of oscillators to the classical
theory of dynamical critical phenomena \cite{four,five} have been made, 
it is interesting to see if the great
deal of results obtained in the past for the finite dimensional  
$XY$ model in the presence of disorder, could possibly relate to the 
problem of arrays of dissipatively coupled, short
ranged oscillators with a natural intrinsic frequency spread. 
We refer to this last problem as the finite dimensional Kuramoto
model or lattice oscillator problem \cite{one}.
Obviously the simplest case one 
may consider is the strictly two dimensional case, where the 
$XY$ type of models are widely studied and where real space renormalization 
group theory has been capable of predicting a great deal of results in the 
past \cite{threee}. 

We will show that the formal analogy between identical, 
dissipatively coupled arrays of oscillators and model A of 
non-equilibrium critical phenomena \cite{fiveb} can be extended, in the two
dimensional case, to the case of non-identical 
oscillators, so that the steady state non-equilibrium properties in the 
system may be studied as a function of the quenched intrinsic frequency
distribution. This will allow us to describe the steady state out of
equilibrium properties of two dimensional oscillator arrays by mean of
classical statistical mechanics. The possibility to consider
non-identical oscillators is essential, in that a finite width of the 
intrinsic frequency distribution is the fundamental parameter one
consider in all Kuramoto type of models. We will also include a physical 
temperature corresponding to white noise in the array and consider 
the case of bimodal frequency distributions, options also 
commonly considered in the context of the mean-field Kuramoto model \cite{zeroc}. 

In this respect we will show that the steady state
properties of a two dimensional array of dissipatively coupled
non-identical oscillators can be related to the equilibrium 
properties of an effective $XY$ model in the presence of random
Dzyaloshinskii-Moriya (DM) interactions. 
The random (DM) interaction effectively induces phase canting between 
neighbouring oscillators, and can be related, as we will show below, to
the quenched distribution of the native intrinsic frequencies. 
It is then argued that 
the steady state non-equilibrium properties of two dimensional arrays of dissipatively 
coupled non-identical oscillators (in the phase approximation) are related to the 
equilibrium properties of the two dimensional $XY$ model in the presence of 
disorder, where again a great deal of real space renormalization group theory 
results are available, even though the effect of quenched
random interactions on the classical properties of the two dimensional
$XY$ model is still a matter of debate within the condensed matter
physics community. 
A similar approach to describe the non-equilibrium
properties of two dimensional superconducting arrays with external currents 
by mean of the equilibrium properties of an effective $XY$ model has
been considered in the past \cite{six}.

Before we begin our analysis, it is important to stress that 
the properties of the Kuramoto model in low dimensions have already been discussed 
in the past \cite{one,two,twob,three}. In particular one could 
argue at this point that in two dimensions no macroscopic mutual
entrainment (MME) is expected, so that collective synchronization
is simply not present. This argument reminds a similar one about the very 
nature of the critical properties of the two dimensional $XY$ model 
\cite{threee,threed}. If magnetization is known not to be expected at finite temperatures, 
the two dimensional $XY$ model is known however to present critical properties of a special 
type (AO). We will show in the following that the dynamical analog of algebraic 
order is present in dissipatively coupled two dimensional arrays of
oscillators (and we might refer to it as algebraic synchronization). 
Dissipatively coupled oscillators in two dimensions are characterized by a 
phase transition of the Berezinskii-Kosterlitz-Thouless (BKT) type, as
we will show in this article, even for non-identical oscillators, namely 
when one consider the non trivial case of distributed intrinsic
frequencies. The main question is how algebraic order (AO) reacts to the 
disruptive effect of the intrinsic frequency distribution, and if 
a steady state algebraic synchronization regime exists in two 
dimensional arrays of oscillators. 

We will address these questions by evaluating the renormalization group 
recursion relations, within the position space Migdal-Kadanoff (MK)
approximation, for the case of an $XY$ model with quenched random 
interactions induced by effective Dzyaloshinskii-Moriya (DM)
interactions. We consider the standard formulation of the $XY$
model within the (MK) approach \cite{sixb}, adopting the
discretization scheme introduced in reference \cite{seven} and 
proposing a novel set of recursion relations that will allow us to 
include both random exchange and random (DM) interactions, 
consistently with the usual methods of real space 
(MK) renormalization group of disordered spin systems \cite{eight}.
We will be considering a renormalization group rescaling length $b=3$, to treat
exchange and random (DM) interactions of random opposite signs on equal
footing \cite{nine}, effectively generalizing the position space
renormalization group (PSRG) recursion relations of the two dimensional
XY model to the case of random exchange and (DM) interactions.
Together with the novel set of (PSRG) recursion relations, we present 
preliminary results obtained by means of an algorithmic
method capable to implement the recursion relations and evaluate 
the corresponding phase diagram, for different model systems, 
corresponding to different forms of disorder, by fixing the proper
choice of the initial conditions in the renormalization group
transformation. Choices we will consider ranges 
from the $XY$ spin glass (XYSG), where randomness is in the exchange
interaction and no (DM) interactions are present, the
two dimensional ferromagnet with Dzyaloshinskii-Moriya interactions
(XYDM) problem, where randomness is in the (DM) term and no random exchange 
interaction term is present, and ultimately the random gauge glass (RGG)
problem, as an important general case where both type of randomness are 
present simultaneously, so that gauge invariance is restored. 
Our MK renormalization 
group approach will be general enough to describe any of the models 
above, by a proper choice of the initial conditions in the
renormalization group flows.

The proper initial condition of primary interest, i.e. the one we 
wish to consider to be able to predict the type of behavior one expects in two
dimensional arrays of dissipatively coupled arrays of non-identical 
oscillators, is, as we now show, of the (XYDM) type, 
even though randomness is the dissipative/exchange coupling term is
also relevant when discussing two dimensional oscillator arrays. 

Our results do not restrain to the oscillator array problem only, but
are general enough and pertain to different type 
of systems that have been considered in the past within the condensed 
matter theory community, and where it is relatively common to consider 
the properties of the $XY$ model in the presence of quenched random
interactions of various types.

The shape of the phase diagram of the two dimensional $XY$ model in 
the presence of quenched random (DM) impurities is the subject of a 
long standing debate \cite{nineteen,twentytwo,nineb,twentyfour} we here 
address and show to relate also to the study of dissipatively 
coupled, two dimensional oscillator arrays.

\section{The Model}

The dynamical behavior of oscillator arrays in the vicinity of a
supercritical Hopf bifurcation can be described in terms of complex 
amplitude equations, related to the amplitude and phase of each
oscillator. In the general case of reactive and dissipative coupling, 
as well as considering the presence of stiffening non-linearities
of the Duffing type, and in the case of an intrinsic frequency distribution
spread within the oscillators, one can write the following equations for the complex
amplitude \cite{ten,eleven}, where, as mentioned above, we add a finite 
temperature due to brownian fluctuations within the array so that $ \langle
\eta_k(t) \eta_{k'}(t') \rangle = 2T \delta_{k,k'} \delta(t-t')$. 
We restrict the range of the interactions to neighbouring 
oscillators on the square lattice.
\begin{eqnarray}
\frac{dA_k}{dt}&=& ( \mu + i \omega_k)A_k-( \gamma+i
 \alpha)|A_k|^2A_k \nonumber \\
&+&( J+i R) \sum_{j \in { \cal L}_k}(A_j-A_k)+\eta_k 
\label{one}
\end{eqnarray}

We will assume that the oscillators are not identical so that quenched 
intrinsic frequencies $\omega_k$ are considered, with a given
(e.g. bimodal) distribution $ P ( \omega)$. The two parameters $ \mu$
and $ \gamma $ are related to the self-sustaining mechanism that drives 
each oscillator; for oscillators of the Van der Pol type we can 
choose $ \mu=\gamma$. The coupling $J$ is the dissipative coupling and 
we here assume to be uniform and short ranged. In what follows we will 
also consider the possibility of a random dissipative/exchange coupling. 
Finally the reactive term $R$, which is related to a mechanical coupling 
between the oscillators, is also short ranged and eventually distributed 
around an average value $R_o$.
We note at this point that the interesting case of reactively coupled 
oscillators has been considered in reference \cite{fourteen}, in the
mean-field approximation. We neglect this last term, and rewrite 
the above equation as

\begin{eqnarray}
 \frac{dr_k}{dt}&=&(\mu-\gamma r_k^2)r_k+J \sum_{j \in { \cal L}_k} 
  r_j \big ( 1-\cos(\theta_j-\theta_k) \big )+\eta^{r}_k \nonumber \\
 \frac{d \theta_k}{dt}&=&(\omega_k - \alpha_k r^2_k)+J \sum_{j \in { \cal L}_k} 
  r_j/r_k \sin(\theta_j-\theta_k)+\eta^{ \theta}_k 
\end{eqnarray}

Including a reactive term means one should include $ R \sum_{j \in { \cal L}_k} 
r_k \sin(\theta_j-\theta_k) $ to the right end side of the first
equation and $R \sum_{j \in { \cal L}_k}  r_j/r_k ( \cos(\theta_j-\theta_k) -1) $
to the right end side of the second.
We consider in what follows the isochronous case, assuming that
the Duffing parameter $ \alpha_k$ can be neglected, as in the original 
formulation of \cite{zerob}. 
The magnitude of the complex amplitude crosses over to one as soon 
as the value of $ \mu$,  
related to the Van
der Pol strength of the autonomous oscillators, is properly tuned.
Assuming the width of the intrinsic frequency distribution to be relatively
high peaked around the average value $ \omega_o$, the above equation reduces 
to the phase 
approximation,  \cite{zerob} and one writes
\begin{equation}
 \frac{d \theta_k}{dt}= \omega_k +J \sum_{j \in { \cal L}_k}
  \sin(\theta_j-\theta_k)+ \eta^{ \theta}_k  
\label{KU}
\end{equation}
where ${ \cal L}_k$ are the neighbouring sites to $k$.
In the simple case of identical oscillators one recover the Langevin equations 
for the $XY$ model. Differently, we are back to the oscillator lattice
problem.
In other terms, as mentioned in the introduction, the dynamical behavior of identical oscillators, 
reduces, up to a global redefinition of the complex amplitudes, to the 
coarsening dynamics of the $XY$ model, so that the above equation
(\ref{one}) reduces to model A of critical dynamical phenomena
\cite{fiveb,twelve,thirteen} and the formal analogy between a critical Hopf
bifurcation and phase transition theory can easily be drawn in this
case. The dynamical behavior of non-identical oscillators is certainly
harder to solve, as usually happens when dealing with models in the
presence of quenched randomness. We are effectively studying
the low-dimensional version of the Kuramoto model, and the first
question is how the random frequency term affects the onset 
of algebraic order (AO) we know to be present in the case of identical oscillators.
We will address these questions in the rest of the paper, showing how the 
non-equilibrium steady state properties of the problem can be related to the
equilibrium properties of an effective $XY$ model with random quenched 
interactions. Before doing so, it is interesting to review how (and if) 
numerical simulations might be useful to answer, at least in part, this 
question before we return to a real space, renormalization group calculation  
in the rest of the paper. 
We clearly expect that disorder, in combination to the finite 
dimensional character of the connectivity, might result in numerical 
simulations to be converging very slowly, a situation we faced and
we will discuss in the following section.

\section{Numerical simulations}

In the current assumption of dissipative coupling and within the phase
approximation we discussed above, we shown that the amplitude equations for the
two-dimensional array reduce to the equations of motion of the $XY$
model, with the inclusion of a quenched random frequency term.
How are the critical properties of the $XY$ model affected by the random
frequency term? We firstly answer this question numerically, by 
means of direct integration methods. We considered a
gaussian form for the intrinsic frequency distribution and solve the 
Langevin dynamics explicitly, via Runge-Kutta methods for a
system of size $N=L \times L$, following
the standard methods discussed in the literature \cite{fifteen,sixteen}. 
\begin{figure}
\includegraphics*[scale=0.8,angle=0]{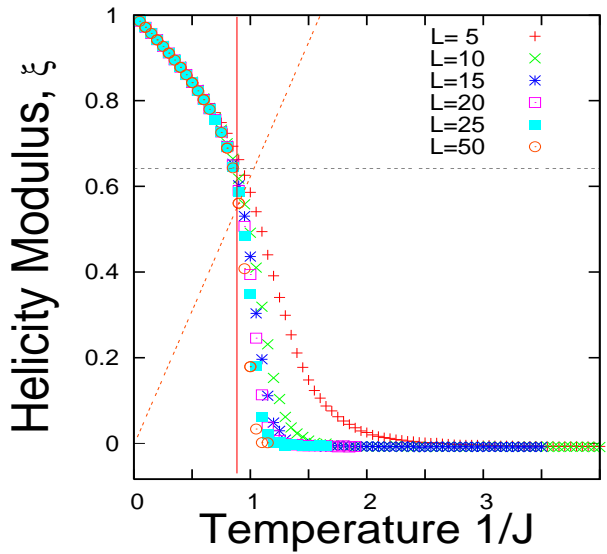}
\caption{Helicity Modulus computed for system sizes $L=5,10,1,20,25,50$ 
in the two dimensional $XY$ model without disorder, by mean of direct 
Runge-Kutta integration methods. The Helicity modulus jump becomes
 steeper for increasing system sizes. We could extrapolate a critical 
transition value of $T_{KT} \simeq 0.89$, in agreement with previous, 
accurate estimates \cite{eighteen}.}
\label{Helix}
\end{figure}
\begin{figure}
\includegraphics*[scale=0.6,angle=270]{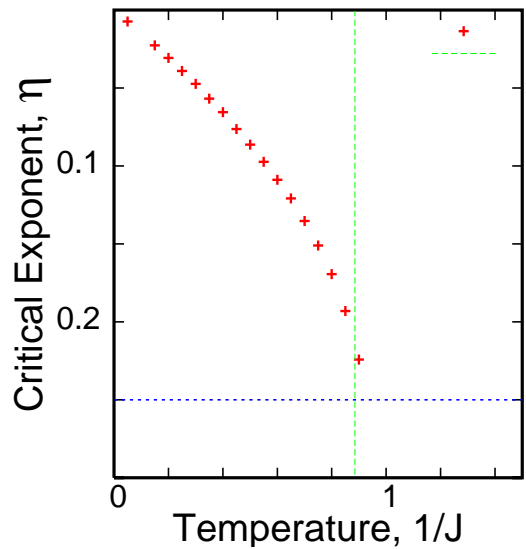}
\caption{Critical exponent $ \eta$. We computed explicitly the 
correlation function at increasing values of the temperature $T<T_KT$, 
observing as expected, algebraic decay. The estimated value of the 
exponent is consistent with the value $ \eta=1/4$, expected at
 $T=T_{KT}$ \cite{threee}.}
\label{Eta}
\end{figure}
We also checked that the 
same algorithm, in the case of mean-field interactions and when random
frequency terms are included, reproduces the results of the Kuramoto model.
We also check that we were able to obtain the expected results in the 
case of identical oscillators in two dimensions. 
Even though the system does not order at finite temperature, phase 
correlation effects are present and are quantified by finite values of the 
helicity modulus \cite{seventeen}, as predicted by the Kosterlitz
Thouless renormalization group theory. The numerical results we here
show are consistent with the expected (BKT) transition
\cite{threee,eighteen}. The algebraic decay of the correlation function 
can be measured and the weak violation of universality critical
phenomena is seen, as expected, with the critical exponent being 
close to the expected value $ \eta=1/4$ at the transition point (see
Fig. \ref{Eta}).
In the presence of disorder, as usually occurs in low dimensional 
systems, things become difficult from the perspective of numerics.
We clearly observed that for each given quenched realization of the intrinsic 
frequency distribution, the system reaches a steady state, i.e., for any
realization of the disorder, we observed that the Runge-Kutta algorithm 
reaches a time independent state. We computed the average helicity
modulus over several realization of disorder, for a given system size
ranging, as in the pure case, from $L=5$ to $L=50$, for two dimensional 
arrays of size $L \times L$. We observed however that the results 
we obtained from numerics in the presence of disorder are not that
clear, the sample-to-sample fluctuations 
becoming very strong as the width of the intrinsic frequency distribution 
grows. For very small system sizes we observe the helicity modulus to 
be finite below the critical point, that decreases with the disorder
strength, as in Fig. \ref{helix_d}. However for systems of size as small as
$N=L \times L=16^2$ it becomes very difficult to have sample-to-sample 
fluctuations under control, so no conclusive results were obtained 
for these system sizes, even after a run of several weeks on a standard 
workstation.
\begin{figure}
\includegraphics*[scale=0.8,angle=0]{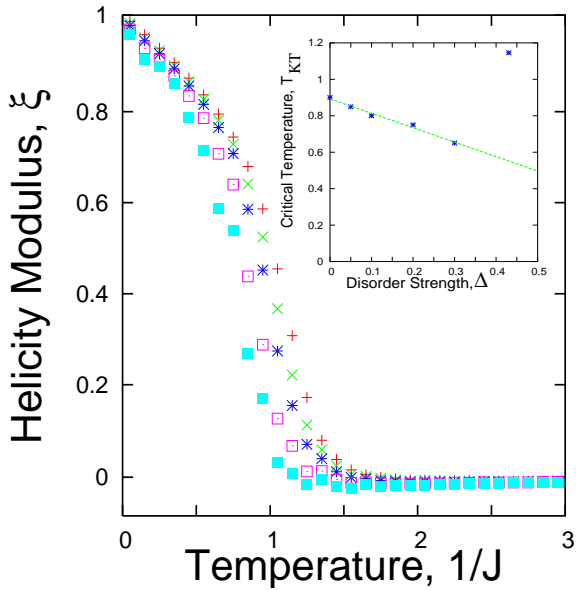}
\caption{Helicity Modulus computed for system size $N=16 \times 16$.
For this small system size we were able to control sample-to-sample 
fluctuations and estimate the helicity modulus for increasing values 
of the disorder strength, $\Delta=0.05,0.1,0.2,0.3$, defined as the 
width of the gaussian distribution of the intrinsic frequency 
distribution $P(\omega)$.}
\label{helix_d}
\end{figure}
Differently, in the mean field case, where each oscillator is coupled to
all others with equal strength, we were able to control 
sample-to-sample fluctuations and an average value of the Kuramoto order 
parameter was easily obtained, consistently with theoretical predictions, up to
relatively large system sizes $N =10^4$. 
To conclude this section on the direct integration approach of
(\ref{KU}), we want to comment further the results we obtained in
the presence of disorder. Our findings seems 
to indicate that the location of the (BKT) transition decreases for increasing
values of the disorder strength, as in the inset. On the other side, in the low
temperature region and for system sizes $L \ge 32$, we do not see the
helicity modulus to reach a constant finite value as for the small
system $L=16$ analyzed in Fig. \ref{helix_d}. In other terms, if algebraic order 
is seen to occur at finite temperatures, for finite, small values of the 
intrinsic frequency width distribution, numerics cannot state in a 
clear manner whether such transition extends all the way down to zero 
temperatures. The helicity modulus sample-to-sample fluctuations diverge 
in the low temperature region, so unique conclusions on the form of the 
phase diagram, in the space of temperature and disorder strength, cannot 
be reached by mean of direct integration methods. We believe that the 
difficulties we just discussed do not really depend on the specific 
method we used, but simply, as often occurs in low dimensional systems,
the presence of disorder makes numerical simulations a very hard task. 

In this section we have seen how direct integration methods cannot state 
in a clear manner how algebraic order is affected by the presence of
random quenched intrinsic frequencies. For this reason we will address 
the same question within real space renormalization group theory in what 
follows.

\section{Real Space Renormalization Group Theory}

The effect of quenched random interactions on the Kosterlitz-Thouless
type transition is an old and interesting problem that applies to a
variety of different physical problems. If one normally
expects that infinitesimal bond randomness do not affect the transition, it has not 
been clarified in a conclusive way how other types of quenched random interactions, namely site
disorder and/or random interactions of the (DM) type affects algebraic 
order. Real space renormalization group considerations suggests that at
finite temperature the (BKT) transition is not destroyed \cite{nineteen}. 
However the effect of quenched random (DM) impurities on the low temperature 
behavior of the system is usually expected to destroy (BKT) order, and 
a reentrant phase diagram has been predicted long ago, 
\cite{nineteen,twenty,twentyfourb}. On the other side, the validity of such
results have been questioned in the recent past
\cite{nineb,twentyeight}, and a Migdal-Kadanoff set of recursion 
relations have been considered \cite{seven}, 
possibly suggesting that algebraic order is stable against disorder 
at finite, small temperatures so that no reentrance would be present. 
Before returning to this point, we want to stress why this question is 
also relevant when studying the effect of a random intrinsic frequency
spread in the two dimensional array of dissipatively coupled
oscillators we consider in this work. 
In the first part of this section we show that disorder in the intrinsic 
frequency distribution of the two dimensional oscillator array problem is
related to the classical types of quenched random interactions 
that have been introduced in the past in the context of the $XY$ model. 
Once this point is clarified and once we will explain how the non-equilibrium
steady state properties of two dimensional arrays of dissipatively
coupled array of oscillators can be understood studying the
equilibrium properties of an effective $XY$ model in the 
presence of quenched random interactions of the (DM) type, we will
critically reconsider the Migdal Kadanoff position space renormalization 
group calculation of reference \cite{nineb} to
establish/check whether and how random (DM) interactions affects 
algebraic order, proposing a new set of recursion relations to evaluate 
the corresponding  phase diagram. We hence suggest the 
correct recursions that can answer the question of how quenched interactions 
affects (BKT) order according to the (MK) approximation, a method that 
has been shown to describe low dimensional systems with quenched
disorder in a simple and effective way. 
We believe that the phase diagram of the $XY$ model, in the
presence of random (DM) interactions within the MK approximation is 
consistent with the reentrant behavior originally 
predicted within real space renormalization group theory \cite{nineteen}. 
We suggest however that reentrance reduces 
at very low temperatures, so that the phase boundary intercept the zero 
temperature axis at finite values of the disorder strength 
(with vertical slope).
In order to obtain the phase diagram, we write 
generalized recursion relations for a discretized clock state model,
following the work of reference \cite{seven}, including bond
and (DM) randomness in a consistent way, at any finite $Q$ values
of the clock state parameter. Our recursion
relations reduce to the well known random bond Ising model for values of the 
clock state parameter $Q=2$ and reduces to the (MK) recursion relations of 
the pure $XY$ model at large values of the clock 
state parameter $Q$, when no disorder is present. 
Without loss of continuity we are able to interpolate between the 
Random Bond Ising model (RBIM) \cite{twentythreeb}, 
the pure $XY$ model at large values of $Q$ and no disorder as well as 
the general case of both exchange and (DM) random interactions being 
present, ultimately proposing a novel framework to study the random gauge
glass model, where both exchange and (DM) interactions coexists 
and are related by the proper initial condition in the renormalization 
group flows. 

The relevance of these models and of the above questions to the
oscillator array problem can be explained as follows. 
Let us consider the following effective hamiltonian

\begin{equation}
 - \beta { \cal H} = \sum_{ \langle i,j \rangle} J_{ij} \vec{S}_i
  \vec{S}_j + 
\sum_{ \langle i,j \rangle} D_{ij} \hat{z} \cdot \vec{S}_i \times \vec{S}_j
\label{DM}
\end{equation}

In the classical formulation of the two dimensional $XY$ model one 
expands around small values of subsequent phase shifts and then 
imposes the periodicity of the phase variables, leading to the canonical 
Coulomb gas formulation \cite{twentythreec}, so the Kosterlitz 
Thouless renormalization group equations can be derived. Similarly, expanding the 
second term related to the random (DM) interaction, an effective
Coulomb gas with random dipole moments can be obtained \cite{nineteen}.
Following these steps and averaging over the disorder either with the 
replica method or alternatively \cite{nineteen,twentyone} recursion relations 
of the Kosterlitz Thouless type that include the presence of disorder have 
been obtained and a reentrant phase diagram in the space of temperature 
and random strength has been predicted. 

The above hamiltonian, considering the case of no disorder being present in 
the exchange interaction $J_{ij}=J_o$ in eq. (\ref{DM}), usually referred 
as the two dimensional ferromagnet with Dzyaloshinskii-Moriya (XYDM) 
interactions, is characterized by equation of motions that, to leading 
order in the adjacent phase shift $\theta_i-\theta_j$
corresponds the the two dimensional Kuramoto model, as a simple
calculation of the equations of motion reveals (see e.g. eq. (9.7) in
reference  \cite{twentyfourb}). We note that the intrinsic frequencies 
$\omega_i$ are related to the quenched dipole distribution according to 
$\omega_i =\sum_{j\in {\cal L}_j} q_{ij}$, where $q_{ij}$ are 
the effective dipoles related to the original (DM) interaction.

The system of oscillators with distributed, quenched frequencies has 
been related to the case of identical oscillators with an effective 
quenched dipole field, that induces a relative frequency mismatch.
According to the above hamiltonian (\ref{DM}), and assuming no 
disorder present in the exchange interaction, the problem relates, 
as one can check explicitly writing the Langevin equations 
associated with (\ref{DM}), to the lattice oscillator problem 
in two dimensions.

It is then reasonable to ask how to write the proper recursion relations 
within the the Migdal Kadanoff approximation, corresponding to the above 
hamiltonian system (\ref{DM}). 

The equilibrium properties of model (\ref{DM}) relates to the steady
state solution of the two dimensional oscillator model we originally 
intended to study. In the next section we present a study of the 
two dimensional $XY$ model with quenched exchange and/or random (DM)
interactions within the MK approximation. 
Not only our finding will be relevant to the 
oscillator array problem but to many other physical problems 
to which the two-dimensional $XY$ model in the presence 
of disorder of the type dictated by equation (\ref{DM}) is known to relate.

\section {The Migdal Kadanoff method}

In the classical formulation of the pure $XY$ model with the Migdal Kadanoff 
renormalization group scheme \cite{sixb}, renormalization 
group recursion relations for the effective coupling of combined bonds, within
a Fourier mode formulation are considered. 
At low temperatures the approximation recovers effective algebraic order, 
the potential flows to a Villain form \cite{sixc} and the system is 
characterized by an infinite correlation length, even though has vanishing 
magnetization. The only drawback of the (MK) approximation is that after 
a sufficiently large number of renormalization group steps, the recursion
relations flows to a  high temperature disordered sink at any finite 
temperature so that true fixed line behavior is not present in the
strict sense. The number of iterations required for this to occur is 
however diverging below the transition point so that effective algebraic 
order is successfully described by the approximation. 
It is then reasonable to ask 
how to reformulate the classical (MK) approach to the $XY$ model in the 
presence of quenched randomness, due to the overall success of the (MK)
approximation in the context of disordered Ising type of models. 

\subsection{The discrete Clock State}

Meanwhile a second, interesting formulation of the (MK) approximation for 
the two dimensional pure $XY$ model has been suggested in the past \cite{seven}.
Rather than the usual Fourier version of the (MK) recursion relations, 
one considers a discrete clock state model, at integer finite values of the 
clock state parameter $Q$. The above formulation has been given for the 
pure $XY$ model in two dimensions. Attempts in order to include the
presence of (DM) interactions have been made \cite{twentyfive,twentyseven,twentyeight} 
but we are not aware of any study where correct recursion 
relations have been written for this case. 
Another delicate point is how one treats the recursion
relations in the case of disorder (both in the exchange and (DM)
interactions). We treat disorder in the same way it has been considered
in the past for the (RBIM), namely within a binning procedure, avoiding 
to resort to random pool methods \cite{twentyseven,twentyeight}, our method 
being simply a deterministic one (despite we are dealing with
disorder). Our recursion relations and algorithmic method reduce to the 
calculation we did in the past for the (RBIM) for values of the clock state 
parameter $Q=2$. 
The computational effort grows linearly with $Q$.
As we will see for the simple case of the pure model, effective 
algebraic order is observed already for values of 
the clock state parameter as small as $Q=32$. 
In order to treat random exchange and (DM) interactions of both sign 
on equal footing we need to consider a renormalization group rescaling 
length $b=3$. This last choice makes the algorithmic solution of the recursion 
relations in the presence of disorder quite slow, since triple
convolutions have to be considered. A similar issue has been discussed 
already for the simple case of the $Q=2$ (RBIM) \cite{twentythreeb}.

\subsection{The Recursion Relations}

As mentioned above we consider the case of discrete phase angles 
$\theta_i=2 \pi q_i/Q, q_i=0,\cdots,Q-1$. 

The hamiltonian reads
\begin{equation}
 - \beta { \cal H} = \sum_{ \langle i,j \rangle} J_{ij} \cos(\theta_i-\theta_j)
+ \sum_{ \langle i,j \rangle} D_{ij} \sin(\theta_i-\theta_j).
\end{equation}
Under renormalization group transformation, we consider two generalized 
couplings between neighbouring sites $J_{ij}$ and $D_{ij}$ independently. 
The renormalization group recursion relations read

\begin{eqnarray}
J'(q)&=& \frac{1}{2} \log(R(Q,q) R^{ \dagger}(Q,q))-G'(Q) \nonumber \\
D'(q)&=& \log(R(Q,q)/R^{ \dagger}(Q,q)) \nonumber \\
G'(Q)&=& \frac{1}{2Q} \sum_{q=0}^{Q-1}\log(R(Q,q)R^{ \dagger}(Q,q)) 
\label{RR}
\end{eqnarray}
where interaction terms are always considered modulo $Q$, and 
where the captive renormalization group constant imposes the constrain 
$\sum_q J'(Q,q)=0$ and a similar constrain for $D'(Q,q)$ equally applies.
The renormalization group polynomials, for the case of a 
rescaling length factor $b=2$ read:

\begin{eqnarray}
R(Q,q) = \sum_{l=0}^{Q-1} e^{\tilde{J}_{12}(l)+\tilde{J}_{23}(q+l)+\tilde{D}_{12}(l)-\tilde{D}_{23}(q+l)} \nonumber
 \\
R^{ \dagger}(Q,q) = \sum_{l=0}^{Q-1} e^{\tilde{J}_{12}(l)+\tilde{J}_{23}(q+l)-\tilde{D}_{12}(l)+\tilde{D}_{23}(q+l)}
\label{RGpols}
\end{eqnarray}
where the ``bond-moved'' exchange interactions are
\begin{equation}
\tilde{J}_{ij}=\sum_{n=1}^bJ_{i_n j_n},
\end{equation}
together with a similar expression for the (DM) interactions $\tilde{D}_{ij}$.
These recursion relations might seem at first sight rather similar to the ones
discussed in the past \cite{seven}, however a important symmetry property is
included in the above recursion relations, that was not discussed in
previous formulations. A set of recursion relations of the 
(MK) type, where the two interactions were treated separately as above, 
was discussed, but only in the so called harmonic approximation 
\cite{twentyfive,twentyfiveb,twentysix}.

The above recursion relations will be firstly discussed for a
renormalization group rescaling length $b=3$, in the absence of disorder, so 
we can check that the onset of effective algebraic order described 
by the (MK) method is properly recovered within the discrete clock state 
formulation and we can determine the effect of an uniform (DM)
interaction on the pure $XY$ model. We will then consider the 
case of small clock state $Q$ values, and include the presence of
disorder, checking that the above recursions reproduce, as expected,
the phase diagram for the (RBIM) at $Q=2$ precisely, before returning to the 
ultimate issue of large values of $Q$ and disorder being present,
corresponding to the $XYSG$, $XYDM$ and $RGG$ models discussed above.

Attempts in this direction were already considered in \cite{seven,twentyseven} 
except we are now able to incorporate the presence of (DM)
interactions in a consistent way. 
The recursion relations (\ref{RR}), despite their 
simplicity, are the only possible ones that properly reflects 
the symmetries of the original hamiltonian, as a direct analysis
reveals. Note that, as also explicitly stated 
in reference \cite{seven}, the symmetry properties of the generalized 
potential there proposed were lost under renormalization group transformation 
in the case of (DM) interactions being present, something that should 
simply not occur. The fact that the recursion relations 
written in the past were not complete is probably the reason behind the 
erratic behaviour of the renormalization group flows, also reported in 
\cite{twentyeight}, whenever interactions of the (DM) type were
considered. Any conclusion on the shape of phase diagram of the $XYDM$ 
problem based on these recursion relations \cite{nineb} should then 
be taken with care.

Differently, in our method the potential maintains its symmetry 
properties under renormalization group transformation 
(see e.g. Fig. \ref{XYDMpotential}), as it should 
be under any (symmetry) group transformation.
Once the novelty of the above recursions is understood,
we need to discuss how to implement them algorithmically, in the
presence of quenched random interactions. 
The way to treat disorder from the algorithmic point of view is a delicate 
issue we should also discuss. As explained above we use a
$deterministic$ method based on a binning procedure, that is known to 
respect the symmetries of the renormalization group
transformation, as already discussed in the context of the (RBIM), that 
corresponds to $Q=2$ in the above recursions.

\section{Identical oscillators}

We first show the results we obtained for the pure model (and $b=3$), 
at large finite values of $Q$, checking that the results of the 
(MK) approximation in the $XY$ model are properly reproduced within the 
discretization scheme here considered and discussing the role of (DM) 
uniform interaction terms, before returning on the issue of disorder, 
we will be able to include via the proper choice of the initial
conditions of the renormalization group flows.

\begin{figure}
\includegraphics*[scale=0.7,angle=0]{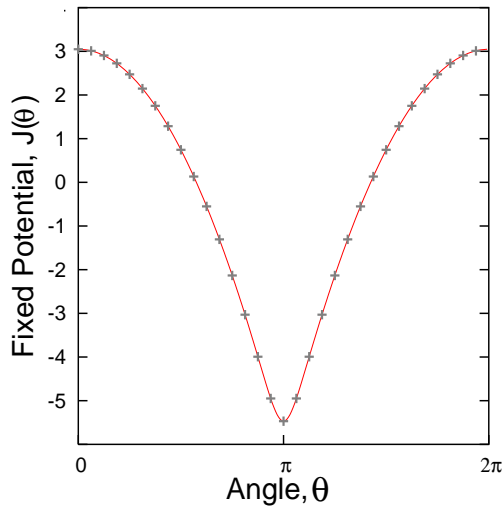}
\caption{The fixed Villain potential observed for the pure $XY$ model, at 
temperature value $T=0.5$ for a renormalization group rescaling 
length $b=3$, in the absence of (DM) interactions. 
Crosses shows the fixed potential in the discrete scheme 
with $Q=32$, while the continuous line shows the results obtained for 
the value of the clock state parameter $Q=512$.}
\label{XYpotential}
\end{figure}

We do not report explicitly the form of the renormalization group 
polynomials for the case of $b=3$, but they clearly include a double sum, 
rather than the single sum in eq.(\ref{RGpols}), as 
usually discussed in the context of the (RBIM) \cite{twentythreeb}, as one 
needs to decimate an even number of sites at each RG step in order to 
treat interactions of both signs in the same manner, so to produce 
a phase diagram that will have the proper symmetry properties. 
Going back to the pure case, when no randomness is present, we expect 
that the location of the Kosterlitz Thouless transition, within the (MK) 
approximation, depends on the decimation parameter $b$, as can be seen 
in Fig. \ref{iter}, showing the number of iterations needed before 
the above recursions (\ref{RR}) flow to the high temperature disordered sink.
Note that in the case of the pure $XY$ model without (uniform) (DM) terms, 
one recovers the usual Villain potential \cite{sixc} for any value 
of the clock state parameter $Q \ge 16$. We show, for the pure
case, both the fixed Villain potential we obtain for $Q=32$ (blue crosses) as 
well as $Q=512$ (continuous red line), in Fig. \ref{XYpotential}. 

\begin{figure}
\includegraphics*[scale=0.7,angle=0]{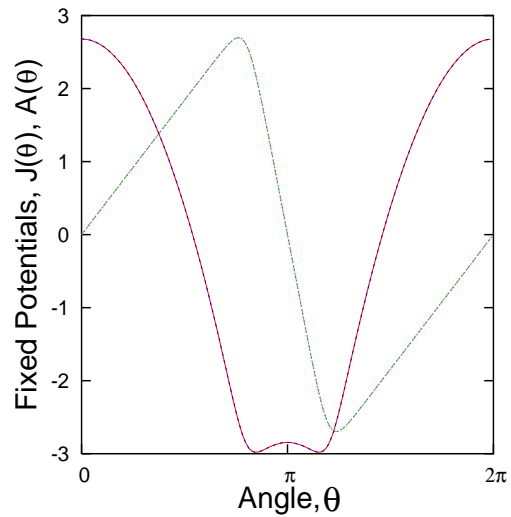}
\caption{The two components of the complex fixed potential, observed 
for the pure $XY$ model, at temperature value $T=0.5$ for a 
renormalization group rescaling length $b=3$, in the presence of uniform 
(DM) interactions, obtained for the clock state parameter value $Q=512$.
Clearly, under renormalization group transformation, the exchange 
interaction is symmetric around $ \pi$, while the (DM) generalized 
interaction is antisymmetric. Fix-line behavior as in the case 
where (DM) interactions are absent is observed.}
\label{XYDMpotential}
\end{figure}

These results shows that the clock state 
discrete approximation is a very robust one, in that effective algebraic 
order is found at relatively small values of the clock state parameter, 
and there is no need to increase the values of $Q$ of the order of
$10^3$, as done in reference \cite{seven}. 
This last remark will be crucial when dealing with
the disordered case, even though the computational effort of the algorithm we
will introduce in the next section ultimately scales in a linear way with
$Q$, while it scales as a cubic power of the number of bins we will use 
to discretize the quenched probability distribution(s) of the exchange 
and (DM) interactions at each renormalization group step.

\begin{figure}
\includegraphics*[scale=0.6,angle=270]{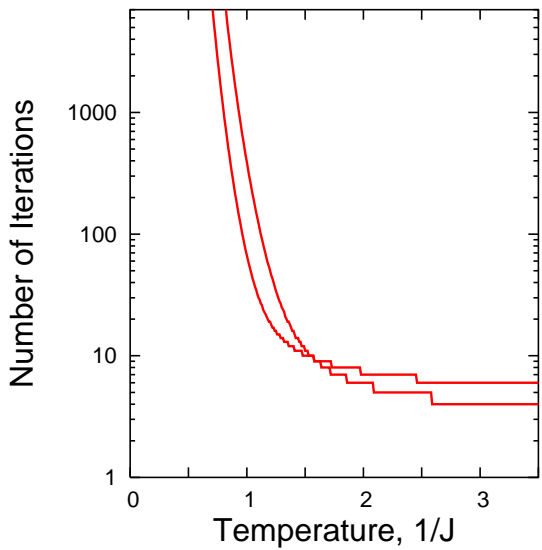}
\caption{The Number of Iterations for the pure $XY$ model 
with $b=2$ and $b=3$. The results shows that effective algebraic order 
is observed in that the increase in the iteration step number is rather 
sharp around the effective critical transition point.The results seems
to suggest that the location of the transition point depend, as occurs 
in Ising type of models within the (MK) approximation, on the
renormalization group rescaling length.}
\label{iter}
\end{figure}

In the presence of (DM) interactions the complex potential
converges to a symmetric and an antisymmetric part ,as in the 
original interaction eq. (\ref{XYDMpotential}). 
Note that the symmetry (antisymmetry) of the two terms in the initial 
hamiltonian are preserved by the above recursion relations (\ref{RR}), 
and that is because we wrote $two$ distinct recursion relations for 
each term, since the renormalization group polynomials have to be 
regarded, in the presence of (DM) interactions,  
effectively as complex quantities. As in the case of the pure $XY$ 
model, effective algebraic order is observed below the 
(BKT) transition point, that do not changes in the presence 
of uniform (DM) interactions, as expected. 

Before we conclude this section on the (MK) clock state approach to the 
pure $XY$ model with uniform (DM) interactions, we want to add 
that, as stressed above, the location of the (BKT) transition do not 
changes importantly for values of the clock state parameter $Q>32$, 
so we can safely consider the disordered case considering finite 
values of $Q$ in this range of values. 

\begin{figure}
\includegraphics*[scale=0.7,angle=0]{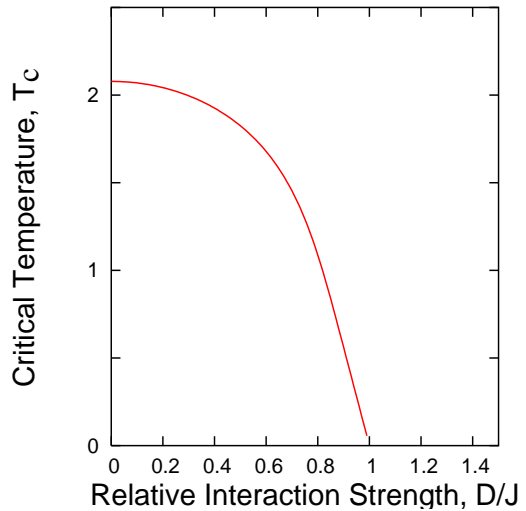}
\caption{The location of the critical transition point for the clock 
state parameter value $Q=4$, when no disorder is present, and where 
uniform (DM) interactions are considered. The $x$-axis shows the 
relative strength of the two interactions, as chosen in the initial
 conditions or the recursions (\ref{RR}).}
\label{DJ}
\end{figure}

For the case of the clock state parameter 
$Q=4$ we observe a ferromagnetic transition point at $T_c \simeq 2.078$ 
that decreases for increasing values of the relative interaction
strength $D/J$, as can be seen in Fig. \ref{DJ}.

\section{The Role of Disorder}

In order to test our recursion relations and algorithmic methods, 
we begin reconsidering the simple (RBIM) case, that is included in the 
above equations at values of the clock state parameter $Q=2,4$, for 
vanishing values of the (DM) interaction strength $D$. 
As expected we recover the reentrant phase diagram that has been 
discussed at length in the past. Clearly, at $Q=2,4$, gauge invariance 
is respected, so we plot the Nishimori line \cite{twentythree} and note 
that, as already found in the past, the phase boundary reaches the highest 
value of the antiferromagnetic bond concentration on this line. 
Note that, at $D=0$ only, the $Q=2$ and $Q=4$ models coincide up to a
redefinition of temperature so we discuss the latter case only.
Interestingly, high precision measures at very low temperatures 
shows that reentrance, that is clearly present as reported 
already \cite{twentythreeb}, diminishes gradually at very low temperatures, 
so that our results are consistent with the expectation that the phase 
boundary is vertical, but at $low$ temperatures only.
Simply this do not occur all the way up to the Nishimori line, according
to our findings. This interesting reentrant behaviour that gradually 
reduces to intercept the zero temperature axis with a vertical slope has also 
been observed recently in the context of (gauge invariant) Gallager 
codes, and we conjecture that the qualitative behaviour of the phase boundary 
we are here discussing applies to all spin systems for higher values of
$Q$, whenever gauge invariance is respected, except the ferromagnetic
phase is replaced by a (BKT) phase for values of the clock state
parameter $Q \ge 16$. 
It is important to mention at this point that any initial condition in 
the (RG) flows for $Q=2,4$, close to the boundary and above the
Nishimori line, will flow to the finite temperature unstable fixed point, 
while any initial condition below the Nishimori line will flow to the 
strong coupling low temperature fixed point already discussed in 
the past \cite{twentythreeb}, a phenomenon known as strong violation 
of universality.

\begin{figure}
\includegraphics*[scale=0.6,angle=0]{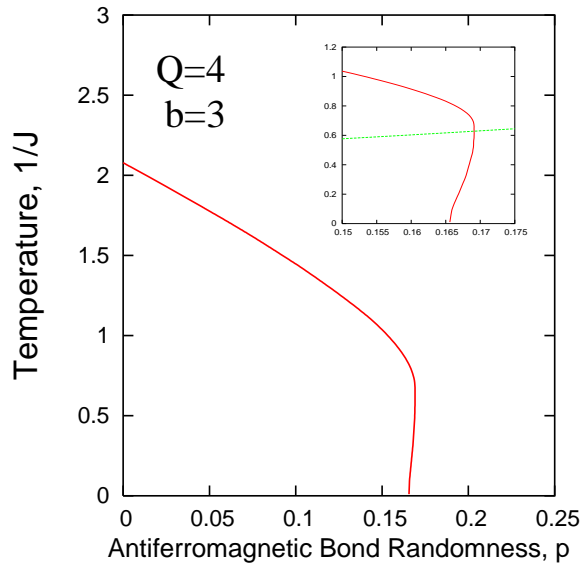}
\caption{
The Phase Digram for the clock state parameter value $Q=4$, without 
(DM) interactions and in the presence of bimodal exchange interactions, 
with an antiferromagnetic bond concentration p. The inset shows the 
reentrant behavior of the phase boundary around the multicritical point.
Gauge invariance is present in the model and so is reentrance.}
\label{PD1}
\end{figure}

\begin{figure}
\includegraphics*[scale=0.8,angle=0]{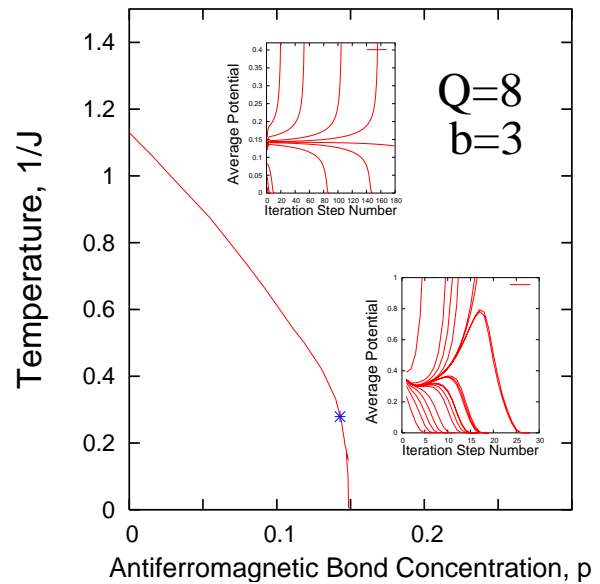}
\caption{The Phase Digram for the clock state parameter value $Q=8$, without 
(DM) interactions and in the presence of bimodal exchange interactions, 
with an antiferromagnetic bond concentration $p$. 
Gauge invariance is not present in this problem, since (DM) interactions
are set to zero. Reentrance is not expected in this case and the
 Nishimori symmetry line is not present. We observe however the presence 
of a multicritical point. Any initial condition above the multicritical 
point and close to the phase boundary, flows to a finite temperature 
unstable fix point, as seen in the inset above, while any initial
 condition below the multicritical point (indicated my the cross) and 
close to the boundary flows to a strong coupling, low temperature fix 
distribution, as in the second inset. The insets plot, as a 
function of the RG iteration number the Average Potential, defined as 
$ \sqrt{ \sum_q |J(Q,q)|^2} $. }

\label{PD2}
\end{figure}

\begin{figure}
\includegraphics*[scale=0.8,angle=0]{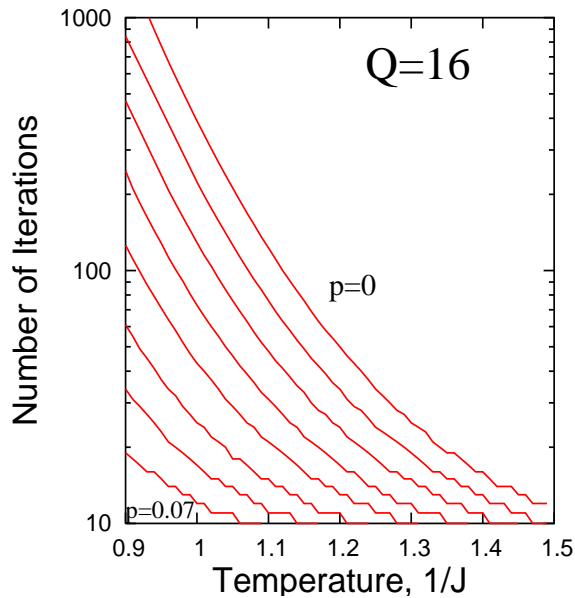}
\caption{ Number of iterations required to flow to the paramagnetic
 sink, as a function of temperature for the 
clock state parameter value $Q=16$ for different values of the 
antiferromagnetic bond concentration $p=0.0, p=0.01, \cdots,0.07$.
The figure indicates that effective algebraic order is observed 
for finite, small values of the disorder strength in the exchange
interaction. From the location of these lines it is possible to 
reconstruct the phase diagram of the $XYSG$ problem at $Q=16$.}

\label{fig10}
\end{figure}

The phase diagram above (Fig. \ref{PD1}) clearly do not relate to 
the original $XY$ model we planning to consider, but we performed 
this calculation to test the algorithm that implements the recursion 
relations in the presence of disorder and we will be 
able to consider large values of $Q$ of the (DM) type at a later stage, 
changing without loss of continuity 
the parameters in the initial condition of the RG flows. Moreover, 
it is quite interesting to study the clock state model 
at small finite values of the parameter $Q$ with disorder, since it is 
a well defined problem that interpolates between the Ising and $XY$ 
type of models. As found in the pure case, when disorder is present 
in the exchange interaction, we observe ferromagnetic order at low 
temperatures for the values $Q=2,4,8$, while effective algebraic order 
appears at values of the clock state parameter $Q \ge 16 $.
This means that the recursion relations (\ref{RR}) flows to a
ferromagnetic fixed point for values of $Q \le 8$, while at $ Q \ge 16$ 
we observe the usual quasi-fixed line behaviour discussed in the context 
of the $XY$ model ( see Fig. \ref{fig10}), and we do not expect things 
to change importantly for higher values 
values of $Q$, as we have seen already for the pure case, meaning that 
the discretization scheme converges rapidly to the original problem 
of continuous degrees of freedom.
For the case of $Q=8$, and when DM interactions are not present 
we computed, for a bimodal exchange interaction, the phase diagram 
in Fig.\ref{PD2}. 
In this case gauge invariance is not present
\cite{twentythree} and reentrance do not occur, as expected. 
On the other side we still 
observe ferromagnetic and paramagnetic order, divided by the phase 
boundary (red thick line in Fig. \ref{PD1}), where a multicritical 
point occur. As in the $Q=2,4$ case, 
any initial condition close to the phase boundary and above the 
multicritical point flows to an unstable finite temperature fix 
point, while any initial condition close to the boundary and 
below the multicritical point flows to a strong coupling zero 
temperature distribution, even though gauge symmetry is not present.
This implies that at small, finite values of the clock state parameter 
the strong violation of universality, discussed in the past for the 
(RBIM) \cite{twentythreeb}, also occurs in the clock state model, 
for values of the clock state parameter $Q \le 8$. 

In the case of larger values of the clock state parameter, e.g. $Q=16$, 
and still at $D=0$, we do not observe the recursion relations (\ref{RR}) 
to flow to a ferromagnetic fix point anymore, but rather quasi-fixed line 
behavior of the (BKT) type appears at finite values of the
antiferromagnetic bond concentration, as seen in Fig. \ref{fig10}. 

We are considering in fact the discrete version of the $XY$ spin glass problem. 
No gauge invariance nor reentrance is expected in this case, and 
we observe effective algebraic order at low values of the bimodal 
exchange interaction, while a paramagnetic phase occurs at higher 
values of disorder and temperature. Clearly no spin-glass order is
observed in two dimension everywhere in the phase diagram.

Global phase diagrams for the 
$XYSG$ and $XYDM$ problems are being evaluated for values of the clock state 
parameter $Q=16,32,64$ and will be presented elsewhere.

We did not consider yet the final case where both randomness 
in the exchange and (DM) interactions are present, even though we are
aware that this was the main question in order to discuss oscillator 
arrays, and leave this question for future work.

We believe we have set up a consistent machinery 
to evaluate the phase diagram in the general case of large values of 
$Q$ and disorder of both types and results in this direction will 
be able to answer on the issue of reentrance for continuous 
spin models, at least within the MK approximation. 

When both disorder in the exchange and (DM) interactions is
present, the renormalization group recursions involve a two dimensional 
probability distribution, so that the binning technique required is slightly 
more complicated than the one considered in the absence of (DM) interactions.
A similar type of situation has been considered 
for the Random Field Random Bond (RFRB) model, where three dimensional 
probability distributions were considered, and it is just 
a technical issue (but rather tedious) to extend the techniques 
developed in that case to the case of random exchange and random (DM) 
interactions, at finite, large values of the clock state parameter $Q$.  

\section{CONCLUSIONS AND FUTURE PERSPECTIVES}

We conclude with a few considerations and predictions on
what we expect to see in the general case of large values of the 
clock state parameter $Q$ and disorder of the (DM) and exchange type, 
according to the preliminary results we obtained so far.

\begin{figure}
\includegraphics*[scale=0.6,angle=0]{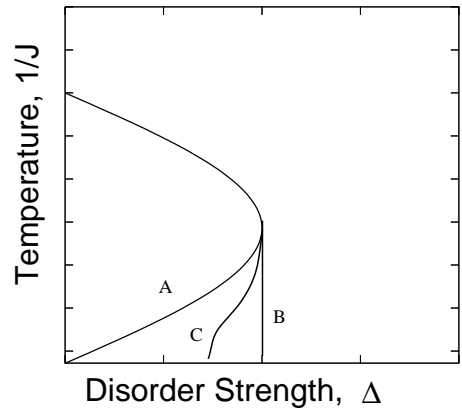}
\caption{ Qualitative behaviour of the phase diagram in the case 
of the two dimensional ferromagnet with (DM) random interactions 
of width $ \Delta$. (A) Real space renormalization \cite{nineteen}, 
suggesting reentrance that extends all the way down to vanishing 
disorder strength. (B) Absence of reentrance and vertical boundary 
at finite values of the disorder strength as suggested, e.g. in
 reference \cite{twentythree}. (C) Reentrant behavior of the 
phase boundary and a (BKT) phase that extend at finite values 
of the disorder strength for low temperature values. }
\label{PD4}
\end{figure}

If both randomness in the exchange and (DM) interaction are considered 
in a way such that gauge invariance is recovered, we expect to observe a (BKT)
phase and reentrant behavior of the phase boundary of the type discussed
above. Note that the exchange randomness is usually considered as 
irrelevant, when (DM) random interactions are present, so similar 
conclusions should apply to the (XYDM) problem.

We expect the phase boundary to reenter below the Nishimori line, 
and finally to intercept the zero temperature axis with vertical slope, 
at finite, $non-vanishing$ values of the disorder strength.
This would imply that (AO) is stable against the intrinsic frequency 
distribution and it is reasonable to ask whether one can observe 
order of the (BKT) in arrays of two dimensional, non-identical
oscillators, where technically one should consider temperature to be
small. Most likely, the reason why previous
calculations within the (MK) approximation have not been able to show the 
reentrance predicted is simply because the recursion
relations considered in \cite{seven,twentyeight} were not taking properly 
into account the (DM) interactions, as stressed above. 
This is likely the reason why the (MK) phase diagram  
discussed in \cite{nineb} is not reentrant, as we also observe for the 
case of $Q=8$ in Fig. (\ref{PD4}) where the (DM) interaction is 
explicitly set to zero and gauge invariance is not present. 
A careful analysis of the recursion relations (\ref{RR}) in the presence 
of (DM) interactions, and the corresponding phase diagram, both 
at small and large values of the clock state parameter $Q$ is in 
progress and will be presented shortly.

The results we expect can be summarised as follows. The phase diagram 
is reentrant as predicted long ago, but reentrance (see curve B in
Fig. \ref{PD4}) reduces and the phase boundary between the (BKT) phase 
and the paramagnetic phase intercept, with vertical slope, the zero 
temperature axis at finite values of the disorder strength. This
intermediate scenario would be consistent both with the reentrant behavior 
predicted by classical renormalization group arguments \cite{nineteen} 
and with the results indicating that the phase boundary intercepts 
the zero temperature axis with vertical slope \cite{twentythree}.

\begin{acknowledgments}
The author wish to thank Prof. David Lowe and Prof. David Saad for useful 
discussions and comments. The author benefited also from brief
but interesting discussions with colleagues during the conference 
in honour of Professor Sherrington, on the occasion of his 65th 
birthday.

The author acknowledge support from EPSRC.
\end{acknowledgments}

\end{document}